\documentclass[11pt,a4paper]{article}
\usepackage{jheppub}
\usepackage[T1]{fontenc}
\usepackage{lmodern}
\usepackage{microtype}
\usepackage{mathtools,bm}
\usepackage{tikz}
\usepackage{xcolor}
\usepackage{orcidlink}
\usetikzlibrary{arrows.meta,decorations.markings,decorations.pathmorphing}
\usepackage[capitalise]{cleveref}
\allowdisplaybreaks[1]
\let\qedthebibliography\thebibliography
\renewcommand{\thebibliography}[1]{%
  \qedthebibliography{#1}%
  \setlength{\itemsep}{2pt plus 0.2ex minus 0.05ex}%
}
\makeatletter
\g@addto@macro\bfseries{\boldmath}
\makeatother
\AtBeginDocument{%
  \raggedbottom
  \setlength{\abovedisplayskip}{8pt plus 2pt minus 2pt}%
  \setlength{\belowdisplayskip}{8pt plus 2pt minus 2pt}%
  \setlength{\abovedisplayshortskip}{2pt plus 1pt}%
  \setlength{\belowdisplayshortskip}{6pt plus 2pt minus 2pt}%
  \setlength{\jot}{2pt}%
}
\hypersetup{
  colorlinks=true,
  linkcolor=blue,
  citecolor=blue,
  urlcolor=blue,
  pdftitle={Hot and dense QED at N3LO: infrared sector},
  pdfauthor={Tyler Gorda, Swastik Majumder}
}
\newcommand{\e}{\mathrm e}
\newcommand{\dd}{\mathrm d}
\newcommand{\ii}{\mathrm i}
\newcommand{\MSbar}{\overline{\mathrm{MS}}}
\newcommand{\order}{\mathcal O}
\newcommand{\Li}{\operatorname{Li}}
\newcommand{\mix}{\mathrm{mix}}
\newcommand{\Pow}{\mathrm{Pow}}
\newcommand{\NLO}{\mathrm{NLO}}
\newcommand{\LO}{\mathrm{LO}}
\newcommand{\hard}{\mathrm{hard}}
\newcommand{\soft}{\mathrm{soft}}
\newcommand{\HTL}{\mathrm{HTL}}

\newcommand{\res}{\mathrm{res}}
\newcommand{\naive}{\mathrm{naive}}
\newcommand{\EFT}{\mathrm{EFT}}
\newcommand{\eps}{\epsilon}
\newcommand{\epsUV}{\epsilon_{\mathrm{UV}}}
\newcommand{\Lam}{\bar{\Lambda}}
\newcommand{\Lamh}{\Lambda_h}
\newcommand{\mE}{m_\mathrm{E}}
\newcommand{\mEd}{m_\mathrm{E,d}}
\newcommand{\gE}{\gamma_\mathrm{E}}

\newcommand{\gHTL}{g_{\mathrm{HTL}}}
\newcommand{\gres}{g_{\HTL}^{\res}}
\newcommand{\gnaive}{g_{\HTL}^{\naive}}
\newcommand{\fHTL}{f_{\mathrm{HTL}}}
\newcommand{\textsumint}{\hbox{$\Sigma$}\!\!\!\!\!\int} 
\newcommand{\sumint}{\hbox{$\sum$}\!\!\!\!\!\!\!\int} 
\renewcommand{\vec}{\bm}
\newcommand{\mubar}{\bar\mu}

\tikzset{
  qed fermion lr/.style={
    draw=black,
    line width=0.75pt,
    postaction={decorate},
    decoration={
      markings,
      mark=at position 0.06 with
        {\arrow{Stealth[length=1.45mm,width=0.95mm]}},
      mark=at position 0.56 with
        {\arrow{Stealth[length=1.45mm,width=0.95mm]}}
    }
  },
  qed fermion tb/.style={
    draw=black,
    line width=0.75pt,
    postaction={decorate},
    decoration={
      markings,
      mark=at position 0.31 with
        {\arrow{Stealth[length=1.45mm,width=0.95mm]}},
      mark=at position 0.81 with
        {\arrow{Stealth[length=1.45mm,width=0.95mm]}}
    }
  },
  qed fermion right arcs/.style={
    draw=black,
    line width=0.75pt,
    postaction={decorate},
    decoration={
      markings,
      mark=at position 0.23 with
        {\arrow{Stealth[length=1.45mm,width=0.95mm]}},
      mark=at position 0.89 with
        {\arrow{Stealth[length=1.45mm,width=0.95mm]}}
    }
  },
  qed fermion top/.style={
    draw=black,
    line width=0.75pt,
    postaction={decorate},
    decoration={
      markings,
      mark=at position 0.31 with
        {\arrow{Stealth[length=1.45mm,width=0.95mm]}},
      mark=at position 0.81 with
        {\arrow{Stealth[length=1.45mm,width=0.95mm]}}
    }
  },
  qed fermion lower left/.style={
    draw=black,
    line width=0.75pt,
    postaction={decorate},
    decoration={
      markings,
      mark=at position 0.15 with
        {\arrow{Stealth[length=1.45mm,width=0.95mm]}},
      mark=at position 0.65 with
        {\arrow{Stealth[length=1.45mm,width=0.95mm]}}
    }
  },
  qed fermion lower right/.style={
    draw=black,
    line width=0.75pt,
    postaction={decorate},
    decoration={
      markings,
      mark=at position 0.48 with
        {\arrow{Stealth[length=1.45mm,width=0.95mm]}},
      mark=at position 0.98 with
        {\arrow{Stealth[length=1.45mm,width=0.95mm]}}
    }
  },
  qed photon/.style={
    draw=black,
    line width=0.65pt,
    decorate,
    decoration={
      snake,
      amplitude=0.85pt,
      segment length=3.7pt
    }
  },
  qed resummed photon/.style={
    draw=black,
    line width=0.48pt,
    double=white,
    double distance=0.95pt,
    decorate,
    decoration={
      snake,
      amplitude=1.0pt,
      segment length=3.9pt
    }
  }
}

\newcommand{\PowDiagram}{%
\begin{tikzpicture}[
  scale=0.62,
  baseline={(current bounding box.center)}
]
  \draw[qed resummed photon]
    (0.151495,0.480005)
    arc[
      start angle=34.3823,
      end angle=325.6177,
      radius=0.85
    ];

  \draw[qed fermion lr]
    (0.42,0) circle[radius=0.55];
\end{tikzpicture}%
}

\newcommand{\NLODiagramA}{%
\begin{tikzpicture}[
  scale=0.62,
  baseline={(current bounding box.center)}
]
  \draw[qed resummed photon]
    (0.151495,0.480005)
    arc[
      start angle=34.3823,
      end angle=325.6177,
      radius=0.85
    ];

  \draw[qed fermion lr]
    (0.42,0) circle[radius=0.55];

  \draw[qed photon]
    (0.42,0.55) -- (0.42,-0.55);
\end{tikzpicture}%
}

\newcommand{\NLODiagramB}{%
\begin{tikzpicture}[
  scale=0.62,
  baseline={(current bounding box.center)}
]
  \draw[qed resummed photon]
    (0.151495,0.480005)
    arc[
      start angle=34.3823,
      end angle=325.6177,
      radius=0.85
    ];

  \draw[qed fermion right arcs]
    (0.42,0) circle[radius=0.55];

  \draw[qed photon]
    (-0.13,0) -- (0.97,0);
\end{tikzpicture}%
}

\title{Hot and dense QED at \texorpdfstring{N$^3$LO}{N3LO}: infrared sector}
\author[1,2]{Tyler Gorda\,\orcidlink{0000-0003-3469-7574}}
\author[1,2]{and Swastik Majumder\,\orcidlink{0009-0004-3006-2904}}
\affiliation[1]{Center for Cosmology and AstroParticle Physics (CCAPP), The Ohio State University, Columbus, Ohio 43210, USA}
\affiliation[2]{Department of Physics, The Ohio State University, Columbus, Ohio 43210, USA}
\emailAdd{gorda.1@osu.edu}
\emailAdd{majumderswastik09@gmail.com}

\abstract{
We determine the infrared-sensitive contribution to the pressure of hot and dense massless quantum electrodynamics (QED) at $\order(e^6)$ in the electric charge $e$, obtaining a representation valid across the crossover between the thermally dominated and degenerate regimes. 
This contribution arises from a hard-thermal-loop (HTL) resummed soft photon dressed by either the one-loop power correction or the two-loop next-to-leading-order photon self-energy. 
After separating the static and nonstatic sectors, we obtain the static contribution in closed form and perform the nonstatic Matsubara sum analytically in terms of the digamma function, reducing the remaining finite coefficient to a one-dimensional numerical integral. 
The finite-temperature HTL counterterm cancels the factorization divergence before expansion; at strictly zero temperature $T$, this divergence instead cancels against the hard four-loop contribution. 
At vanishing chemical potential, the static sector reproduces the complete known \(\order(e^5)\) pressure. 
Taking the cold limit of the dimensionally regulated, unsubtracted nonstatic contribution reproduces the correct cold mixed result. 
Crucially, at any fixed \(T>0\), the nonstatic result remains analytic in \(e^2\). 
The zero-temperature \(\order(e^6\ln e)\) term in QED arises only in the nonuniform limit \(T/\mE\to0\) where $\mE$ is the electric screening mass, demonstrating that the weak-coupling and zero-temperature limits do not commute. 
}

\begin{document}
\maketitle
\section{Introduction}

\label{sec:strategy}

Perturbative calculations of the thermodynamic pressure in gauge theories involve parametrically separated momentum scales.  
In massless quantum electrodynamics (QED), thermal and dense fermionic fluctuations carry hard momenta of order \(Q_{\hard}\sim\max(\pi T,|\mu|)\) with $T$ the temperature of the medium and $\mu$ the fermionic chemical potential, whereas electric screening introduces the soft scale \(\mE\sim eQ_{\hard}\).  
Photon propagation at the latter scale must be resummed using the hard thermal loop (HTL) resummation scheme~\cite{Braaten:1989mz,Blaizot:2001nr} (or  hard dense loop~\cite{Manuel:1995td,Ipp:2003cj,Ipp:2006ij}).  
As has been shown previously in both QED and Quantum Chromodynamics (QCD) at vanishing temperatures, the pressure receives distinct contributions from these different kinematic regions~\cite{Gorda:2018gpy,Gorda:2021znl,Gorda:2021kme,Gorda:2022zyc,Karkkainen:2025nkz}, which can be computed using the HTL effective field theory (EFT) or by computing corrections within it~\cite{Gorda:2022fci,Gorda:2023zwy,Ekstedt:2023anj,Ekstedt:2023oqb}.
These different contributions have been labeled in Ref.~\cite{Gorda:2021znl} as ``soft'', ``mixed'', and ``hard'' by whether they are characterized by pressure contributions stemming from leading-order (LO) HTL, subleading corrections to HTL, and contributions from the full theory, respectively.

One way to understand this organization is within the language of Ref.~\cite{Kurkela:2016was}, namely one can consider the pressure $p$ computed using all necessary resummations that would cure various infrared (IR) divergences that appear when performing a naive loop expansion.\footnote{We thank Pablo Navarrete for encouraging us to consider this alternative understanding of the kinematic sectors.}
One may also consider an EFT of the IR sector and add and subtract the resummed pressure of this EFT
\begin{equation}
    p^\res = p^\res - p^\res_\EFT + p^\res_\EFT = (p^\naive - p^\naive_\EFT) + p^\res_\EFT.
\end{equation}
The second equality follows from the fact that the EFT correctly captures all of the IR physics, and so expanding both the full theory and the EFT in a naive power series in the coupling will result in precisely the same IR structures that will cancel between the two terms.
The quantity $p_\EFT^\naive$ is effectively a counterterm, and one has a freedom to either pair it with the full theory contribution (thus keeping all of the ``naive'' terms together), or to keep it with the rest of the EFT contributions
\begin{equation}
    p^\res = p^\naive + (p^\res_\EFT - p^\naive_\EFT),
\label{eq:res-naive-split}
\end{equation}
the latter of which matches the organization in Ref.~\cite{Kurkela:2016was} and has the advantage of cleanly generalizing to the case when multiple EFTs are required to handle the IR sector.
In this work, we follow this second option and keep the entire IR sector described by the EFT together.

Returning to QED, since screened HTL photons do not interact \cite{Bellac:2011kqa}, the fully soft sector contributes only a ``free'' contribution at next-to-next-to-leading order in the fine structure constant $\alpha_e$, first computed at $T = 0$ in Ref.~\cite{Freedman:1976dm}.
At next-to-next-to-next-to-leading order\footnote{In this work, we define a full order in the expansion to include all terms up to the next integer power of $\alpha_e$.} (N$^3$LO) the pressure takes the form~\cite{Gorda:2021kme,Gorda:2022zyc}
\begin{equation}
  p^{\text{N}^3\LO}=p_{\hard}+p_{\mix}+p_{\soft},
  \qquad p_{\soft}=0\quad\text{in QED}.
  \label{eq:pressure-regions}
\end{equation}
The hard contribution is the naive full-theory contribution in \cref{eq:res-naive-split}, while the mixed contribution contains a soft resummed photon coupled to a hard subdiagram, minus the corresponding HTL counterterm. 

At N$^3$LO, the mixed sector is generated by an HTL-resummed photon dressed by either a one-loop power correction or a two-loop, next-to-leading-order (NLO) photon self-energy~\cite{Manuel:2016wqs,Gorda:2022fci}.
These hard-soft terms contain a factorization ultraviolet (UV) divergence that cancels against the HTL counterterm.
At zero temperature, this counterterm becomes scaleless and vanishes in dimensional regularization \cite{Gorda:2021kme}.
In this case, the UV divergence of the mixed sector cancels with an IR divergence in the hard sector, which leads to a nonanalytic logarithm of the coupling from the ratio of the scales of the hard and HTL theories.  
At nonzero temperature, however, the spacing between the discrete Matsubara modes provides an additional scale, and the appearance of this logarithmic behavior depends on the hierarchy between $T$ and the screening scale $\mE$.

The zero-temperature mixed and partially hard contributions were obtained in Ref.~\cite{Gorda:2022zyc}, while the photon self-energies required here were derived from Ref.~\cite{Gorda:2022fci}.  
The latter calculation organizes the soft-momentum expansion into angular structures and radial moments of the fermion distributions. 
In this work we evaluate the corresponding mixed contribution to the pressure at nonzero temperature and chemical potential.
We adopt the same decomposition for $N_f$ massless leptons with a common chemical potential $\mu$.

The paper is organized as follows. 
In \cref{sec:setup}, we discuss our notational conventions as well as the form of the self-energies that enter our expressions.
In \cref{sec:matching} we define the terms that contribute to the mixed sector at finite-$T$ and compute them. 
We discuss the sum over the nonzero Matsubara modes in \cref{sec:nonzero-modes} and separately treat the zero mode in \cref{sec:zero-mode}.
In \cref{sec:results} we present our results, and discuss the various limiting cases of our expressions, comparing to the existing literature.
We conclude with a summary of our results, as well as a brief discussion of what additional complications are  expected when evaluating the mixed sector of QCD at finite temperatures.

\section{Setup and self-energy structure}\label{sec:setup}

We consider a hot and dense QED plasma at temperature $T$ and electron chemical potential $\mu$.
We introduce the dimensionless ratio
$\mubar \coloneqq \mu/(2\pi T)$, which measures the chemical potential in
units of the nonzero bosonic Matsubara spacing.  
In our computation, we work in Euclidean signature and dimensional regularization, in $D \coloneqq d+1$ spacetime dimensions, with $d \coloneqq 3 - 2\epsilon$ spatial dimensions.
At \(d=3\), the leading-order electric screening mass is
\begin{equation}
  \mE^2 \coloneqq N_fe^2(\Lam)
  \left(\frac{T^2}{3}+\frac{\mu^2}{\pi^2}\right).
  \label{eq:screening-mass}
\end{equation}
We define the momentum of the soft photon as $K_n \coloneqq (\omega_n,\vec{k})$ with $\omega_n \coloneqq 2\pi nT$, and we write $k \coloneqq |\vec{k}|$ so that $K_n^2 = \omega_n^2+k^2$.
The bosonic sum-integral that we need when integrating over the momentum of the soft photon is
\begin{equation}
  \sumint_K
  \coloneqq
  \left(\frac{\e^{\gE}\Lamh^2}{4\pi}\right)^{\eps}
  T\sum_{n\in\mathbb Z}
  \int\frac{\dd^d k}{(2\pi)^d}.
  \label{eq:sumintegral}
\end{equation}
As this is an integral in the soft theory, the $\MSbar$ scale $\Lamh$ appears in the factor in front, rather than the hard scale $\Lam$ in the underlying theory.  
For any $n\ne0$, we define the continuous variable $x$, related to the polar
angle in the full $D$-dimensional space, as
\begin{equation}
  x\coloneqq\frac{k}{\lvert\omega_n\rvert},\qquad
  x\in[0,\infty).
  \label{eq:x-LE}
\end{equation}
In what follows, we will work extensively with the $(\omega_n, x)$ variables in place of the original $(\omega_n, k)$ ones.
Explicitly, we can rewrite the sum over the \emph{nonzero} Matsubara modes (which we denote as $\textsumint\,\!'$) as 
\begin{equation}
   \sumint_{K}' = 2   \left(\frac{\e^{\gE}\Lamh^2}{4\pi T^2}\right)^{\eps}\!\!
  T^{4}\sum_{n = 1}^\infty
  n^d \! \int_0^\infty \dd x \, x^{d-1} \! \int_{S^{d-1}} \!\!\!\! \dd \Omega_{d},
\label{eq:nonzero_mode_change_of_variables}
\end{equation}
over any (sum)integrand that is symmetric under $n \mapsto -n$.
In the above expression, $\dd \Omega_d$ denotes the measure on $S^{d-1}$.

For a soft external photon momentum, the one- and two-loop photon self-energies may be expanded as \cite{Manuel:2016wqs,Carignano:2017ovz,Carignano:2019ofj,Gorda:2022fci}
\begin{align}
  \Pi_{1,I}(K)
    &=\Pi_I^{\LO}(x)+K^2\Pi_I^{\Pow}(x)+\order(K^4),\nonumber\\
  \Pi_{2,I}(K)
    &=\Pi_I^{\NLO}(x)+\order(K^2),
  \label{eq:selfenergy-expansion}
\end{align}
with $I \in \{\mathrm{T}, \mathrm{L}\}$ denoting the polarization.
Note that the various HTL self-energies in Eq.~\eqref{eq:selfenergy-expansion} are functions of $x$ only, which is why the change of variables in Eq.~\eqref{eq:nonzero_mode_change_of_variables} is so useful in our case.
The numbers of transverse and longitudinal modes are $g_{\mathrm T} \coloneqq d-1$ and $g_{\mathrm L} \coloneqq 1$, respectively.
The radial--angular representation of the Euclidean self-energies, the
finite-temperature radial moments, and the complete \(\order(\eps)\)
transverse and longitudinal expressions are collected in \cref{app:selfenergies}.

\section{Computing the mixed contribution}
\label{sec:matching}

Following Refs.~\cite{Gorda:2021kme,Gorda:2022zyc} and Ref.~\cite{Kurkela:2016was}, the mixed QED
pressure is decomposed as
\begin{align}
  p_{\mix}
  &=
  p_{\mix}^{\Pow}  
  +
  p_{\mix}^{\NLO},
  \\[2mm]
  p_{\mix}^{\Pow}
  &\coloneqq
  \vcenter{\hbox{\PowDiagram}} - p_\naive^\Pow,
  \\[2mm]
  p_{\mix}^{\NLO}
  &\coloneqq
  \vcenter{\hbox{\NLODiagramA}}
  \;+\;
  \vcenter{\hbox{\NLODiagramB}}- p_\naive^\NLO,
\end{align}
where the double wavy line denotes an HTL-resummed photon propagator, solid lines are fermions, and the single wavy line denotes a bare photon. 
The naive terms in the above expressions are defined to be expanded versions of the corresponding HTL diagrams, where the HTL-resummed photon propagator has been expanded through one LO insertion for the NLO correction and through two LO insertions for the power correction.
A sum over the direction of fermion flow is understood.
Since the two self-energy corrections for a soft photon have been previously computed, what remains to be performed is the sum-integral over the soft photon.

Defining the HTL resummed and free photon propagators as 
\begin{equation}
  G_I(K)\coloneqq\frac{1}{K^2+\Pi_I^{\LO}(K)},
  \qquad
  \Delta_0(K)\coloneqq\frac{1}{K^2},
  \label{eq:propagators}
\end{equation}
the explicit expressions for the components are
\begin{align}
  p_{\mix}^{\NLO}
  ={}&-\frac{1}{2}\sumint_K\sum_{I=\mathrm T,\mathrm L}g_I\,
    \Pi_I^{\NLO} [G_I-\Delta_0 +\Delta_0\Pi_I^{\LO}\Delta_0], \\
 p_{\mix}^{\Pow}
  ={}&-\frac12\sumint_K\sum_{I=\mathrm T,\mathrm L}g_I\,
    K^2\Pi_I^{\Pow} [G_I-\Delta_0+\Delta_0\Pi_I^{\LO}\Delta_0 - \Delta_0\Pi_I^{\LO}\Delta_0\Pi_I^{\LO}\Delta_0],
\end{align}
where we have suppressed the dependence on $K$ from the self-energies and propagators.
The one-loop power correction is renormalized together with the lower-order
HTL ring. Equivalently, one may use the bare self-energies and include the
coupling counterterm of that ring; the expressions below use the renormalized
self-energies in \cref{app:selfenergies}.
These two terms can be summed to give the compact expression \cite{Gorda:2021kme}
\begin{equation}
  p_{\mix}
  =\frac{1}{2}\sumint_K\sum_{I=\mathrm T,\mathrm L}g_I [\Pi_I^{\NLO}\Pi_I^{\LO}
    -\Pi_I^{\Pow}(\Pi_I^{\LO})^2]
  \left[ \frac{1}
       {K^2(K^2+\Pi_I^{\LO})}  - \frac{1}{(K^2)^2}\right].
\label{eq:pmix-common}
\end{equation}
We define the last term in \cref{eq:pmix-common} to be the $\order(e^6)$ naive term
\begin{equation}
  p_{\mix}^\naive\coloneqq\frac12\sumint_K\sum_I g_I
  \frac{\Pi_I^{\NLO}\Pi_I^{\LO}-\Pi_I^{\Pow}(\Pi_I^{\LO})^2}{(K^2)^2}, 
  \label{eq:hard-four-loop-ir}
\end{equation}
while we label the first term the resummed mixed term. 
Thus the convention used here is
\begin{equation}
  p_{\mix}=p^{\res}_\mix-p_{\mix}^\naive,\qquad
  p_{\hard}^{(4)}=p_{\mathrm{QED}}^{(4), \naive}.
  \label{eq:hard-overlap-matching}
\end{equation}
The alternative convention combines $p_{\mix}^{\res}$ with
$p_{\mathrm{QED}}^{(4),\naive}-p_{\mix}^\naive$; the sum is identical before
expanding in $\eps$. We can now separately compute the ${n=0}$ and
${n\ne0}$ Matsubara modes, which we do in the following two subsections.

\subsection{Contribution from the nonzero Matsubara modes}
\label{sec:nonzero-modes}

For the nonzero modes, we perform the change of variables given in Eq.~\eqref{eq:nonzero_mode_change_of_variables}.
Since the HTL self-energies are functions of $x = k/|\omega_n|$ only, we can write
\begin{equation}
\begin{aligned}
  \left.p_{\mix}\right|_{n\ne0}
  ={}&\left(\frac{\e^{\gE}\Lamh^2}{4\pi T^2} \right)^\eps
  \frac{\Omega_d}{(2\pi)^4}
  \sum_{I=\mathrm T,\mathrm L}g_I
  \int_0^\infty\dd x\,\frac{x^{d-1}}{(1+x^2)^2}
  \left[\Pi_I^{\NLO}\Pi_I^{\LO}
    -\Pi_I^{\Pow} (\Pi_I^{\LO})^2 \right]
  \\[-0.2em]
  &\times
  \sum_{n=1}^{\infty}
  \left[
    \frac{n^{d-2}}{n^2 + \tau^2_I(x)} - 
     n^{d-4}
   \right]
\end{aligned}
  \label{eq:nonstatic-master}
\end{equation}
with the dimensionless screening argument $\tau_I^2(x)$ defined as
\begin{equation}
    \tau^2_I(x) \coloneqq \frac{\Pi_I^{\LO}(x)}{(2 \pi T)^2(1+x^2)},
    \label{eq:screening-argument}
\end{equation}
and $\Omega_d$ the area of $S^{d-1}$,
\begin{equation}
  \Omega_d=\frac{2\pi^{d/2}}{\Gamma(d/2)}.
  \label{eq:omega-d}
\end{equation}
For any $T>0$, the difference of the two sums in
\cref{eq:nonstatic-master} is ultraviolet finite. Define
\begin{equation}
  H_d(\tau)\coloneqq\sum_{n=1}^{\infty}\frac{n^{d-2}}{n^2+\tau^2}.
  \label{eq:Hd-definition}
\end{equation}
Although $H_d(\tau)$ and $H_d(0)$ separately contain the same pole,
their difference can be evaluated directly at $d=3$:
\begin{align}
  \mathcal D(\tau)
  &\coloneqq\lim_{\eps\to0}\bigl[H_{3-2\eps}(\tau)-H_{3-2\eps}(0)\bigr]
  \nonumber\\
  &=-\sum_{n=1}^{\infty}\frac{\tau^2}{n(n^2+\tau^2)}
   =-\frac12\aleph(1+\ii\tau)-\gE.
  \label{eq:Hd-eval}
\end{align}
Here we use the notation of Ref.~\cite{Vuorinen:2003fs},
$\aleph(z)\coloneqq\psi(z)+\psi(z^*)$, with
$\psi(z)\coloneqq\Gamma'(z)/\Gamma(z)$ the digamma function.
The pole cancels before the $\eps$ expansion, so the
$\order(\eps)$ self-energies, the dimensional correction to $\mEd$, and
the $\order(\eps)$ angular measure cancel in this difference at $T>0$.
Appendix~\ref{app:resummed-cold} gives the dimensionally regulated
representation used for the strict $T=0$ calculation.
For general $T$ and $\mu$, the remaining $x$ integral must be performed numerically.
\subsection{Contribution from the zero Matsubara mode}
\label{sec:zero-mode}

The static limits of all required self-energies are collected in \cref{eq:static-checks}.  
In particular, the transverse contribution vanishes, while the longitudinal part of the matched pressure becomes
\begin{align}
  \left.p_{\mix}\right|_{n=0}
  &=\frac{T}{2}
  \left[
    \mE^2\Pi_{\mathrm L}^{\NLO}(0,\vec{k})
    -
    \mE^4\Pi_{\mathrm L}^{\Pow}(0,\vec{k})
  \right]
  \int_{\vec{k}}
  \frac{1}{k^2(k^2+\mE^2)}.
  \label{eq:static-before-integral}
\end{align}
The additional $-1/k^4$ subtraction in \cref{eq:pmix-common} is a
scaleless power integral and vanishes in dimensional regularization.
With \(\int_{\vec{k}}\coloneqq(\e^{\gE}\Lamh^2/4\pi)^\eps
\int\dd^d k/(2\pi)^d\), the remaining radial integral is
\begin{equation}
 \left.\int_{\vec{k}}\frac{1}{k^2(k^2+\mE^2)}\right|_{d=3}
  =
  \frac{1}{4\pi\mE},
  \label{eq:static-integral}
\end{equation}
so that
\begin{equation}
  \left.p_{\mix}\right|_{n=0}
  =
  -\frac{e^2 \mE^3 T}{192 \pi^3} \left[9 + 2 N_f (\mathcal{I}_T - 3) \right],
  \label{eq:static-final}
\end{equation}
after inserting the static limits of the self-energies from \cref{eq:static-checks}. 
Here the thermal function which appears is given by
\begin{equation}
  \mathcal I_T
  \coloneqq 2\Biggl[1+\ln\!\left(\frac{\Lam}{4 \pi T}\right)
  -\frac{1}{2}\aleph \biggl( \frac{1}{2} - \ii \mubar \biggr) \Biggr].
  \label{eq:IT-result-explicit}
\end{equation}
We see the zero-mode contribution is finite in dimensional regularization. 
Since $\mE\sim e$, it contributes at order $e^5$ in weak-coupling counting.
We note that this contribution can also be viewed as the NLO correction to the effective (Debye) mass parameter in the dimensionally reduced electrostatic QED theory.

\section{Results}
\label{sec:results}

The mixed contribution is the sum of the static mode and the nonstatic sum.  
With the overall normalization of the latter factored out, we can define a function
\begin{equation}
  \left.p_{\mix}\right|_{n\ne0}
  =
  \frac{e^2\mE^4}{(4\pi)^4}\,\gHTL\!\left(
    \frac{T}{\mE},\frac{\mu}{T},
    \frac{\Lam}{T},N_f
  \right),
  \label{eq:gHTL-definition}
\end{equation}
such that the complete mixed result is 
\begin{equation}
  p_{\mix}
  =
  -\frac{e^2\mE^3T}{192\pi^3}
    \left[9+2N_f(\mathcal I_T-3)\right]
  +\frac{e^2\mE^4}{(4\pi)^4}\,\gHTL
,
  \label{eq:result-final}
\end{equation}
where the thermal function appearing in the static term is given by \cref{eq:IT-result-explicit}.
The fully reduced representation of $\gHTL$, with its analytic angular
contribution and explicit digamma kernels, is given in
\cref{eq:gHTL-explicit}.
It follows from the subtracted angular average in
\cref{eq:nonstatic-master,eq:Hd-eval}. Here $\mE$ is the $d=3$ mass in
\cref{eq:screening-mass}. The coefficient $\gHTL$ is dimensionless, with its
dependence on the hard renormalization scale and $\mu/T$ displayed explicitly,
and is independent of $\Lamh$ at $T>0$.

For the angular representation, set $x=\tan\phi$ with
$0\leq\phi\leq\pi/2$. Since
$x^2\dd x/(1+x^2)^2=\sin^2\phi\,\dd\phi$, we may write
\begin{align}
  \gHTL&=\int_0^{\pi/2}\dd\phi\,G(\phi),
  \label{eq:gHTL-angular-area}\\
  G(\phi)&\coloneqq\frac{64\pi\sin^2\phi}{e^2\mE^4}
  \left.
  \sum_{I=\mathrm T,\mathrm L}g_{I,0}
  \left[\Pi_I^{\NLO}\Pi_I^{\LO}
    -\Pi_I^{\Pow}(\Pi_I^{\LO})^2\right]_{\eps=0}
  \mathcal D(\tau_{I,0})
  \right|_{x=\tan\phi},
  \label{eq:Gphi-definition}
\end{align}
where $g_{\mathrm T,0}=2$, $g_{\mathrm L,0}=1$, and
$\tau_{I,0}$ is given in \cref{eq:finite-screening-arguments}. Its total,
transverse, and longitudinal contributions are shown in the right
panel of \cref{fig:digamma-crossover}.
Note that $\gHTL$ plays a role analogous to the $\fHTL$ from
Ref.~\cite{Kurkela:2016was}. In the convention used here, the complete
$\order(e^6)$ $p_\mix^\naive$ is subtracted from the mixed contribution and remains
in the infrared behavior of the naive hard sector in \cref{eq:hard-overlap-matching}.

\subsection{Limiting cases}
\label{sec:limiting-cases}

Our result is constrained by two complementary limits. At
\(\mu=0\), the static sector must reproduce the known \(\order(e^5)\)
pressure of hot QED. At strictly zero temperature, the dimensionally
regulated sum must reconstruct the continuous energy integral and
recover the mixed contribution of Ref.~\cite{Gorda:2022zyc}, with its finite
NLO mixed term corrected.\footnote{There are some errors in the results of Ref.~\cite{Gorda:2022zyc} arising from missing terms from expanding $\mEd$ out to $\order(\eps)$ in the NLO self-energies.} 
The finite-temperature interpolation is controlled by the digamma function
appearing in the exact subtracted sum in \cref{eq:Hd-eval}; its
$\eps$-expanded cold limit must be distinguished from setting $T=0$ before
the dimensional expansion.

At zero chemical potential,
\begin{equation}
  \mE^2=\frac{N_fe^2T^2}{3},
  \quad\!\!
  \left.\mathcal I_T\right|_{\mu=0}
  =
  2\left[
    1+\ln\biggl(\frac{\mathrm{e}^{\gE}\Lam}{\pi T}\biggr)
  \right].
  \label{eq:hot-static-inputs}
\end{equation}
Inserting these expressions into the static result
\cref{eq:static-final} gives
\begin{align}
  \left.\frac{p_{\mix}}{T^4}\right|_{\mu=0,\,\order(e^5)}
  ={}&
  -\frac{e^5N_f^{3/2}}{64\sqrt{3}\,\pi^3}
 -\frac{e^5N_f^{5/2}}{144\sqrt{3}\,\pi^3}
  \left[
    \ln\left(\frac{\Lam}{\pi T}\right)
    +\gE-\frac12
  \right],
  \label{eq:hot-e5-match}
\end{align}
which agrees
with Refs.~\cite{Coriano:1994re,Zhai:1995ac}.

\begin{figure}[t]
  \centering
  \includegraphics[width=0.98\linewidth]
    {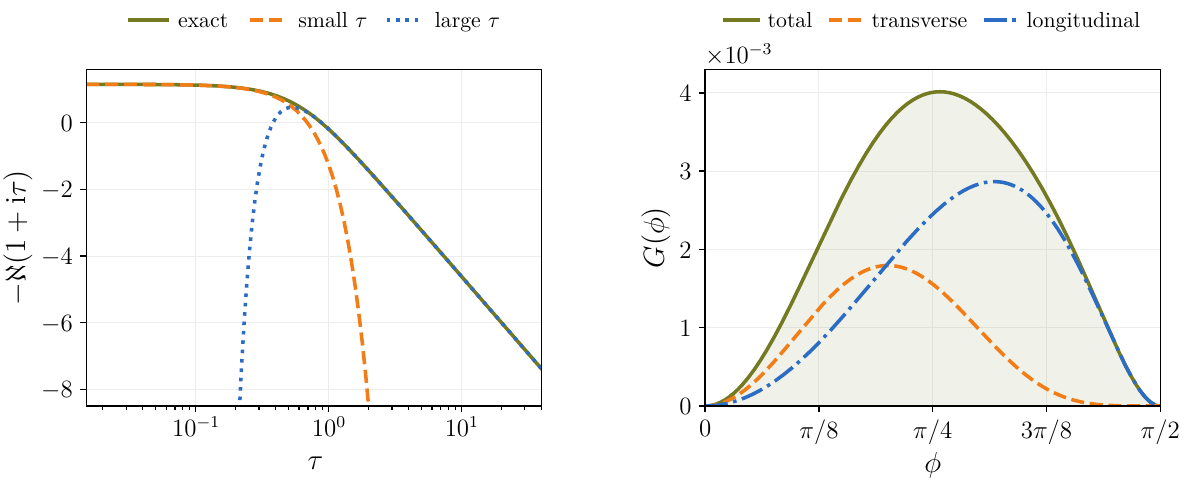}
  \caption{(Left) The function
  \(-\aleph(1+\ii\tau)=-2\operatorname{Re}\psi(1+\ii\tau)\) (solid), its small-\(\tau\) expansion
  through \(\tau^2\) (dashed), and its large-\(\tau\) expansion through
  \(\tau^{-4}\) (dotted). Both expansions are plotted over the full
  interval; the vertical range resolves the crossover.
  (Right) The angular integrand $G(\phi)$ defined in
  \cref{eq:Gphi-definition}. The area under the curve gives $\gHTL$ defined in~\cref{eq:gHTL-explicit}. The parameters used are $T=0.23~\mathrm{GeV}$, $\mu=0.79~\mathrm{GeV}$,
  $\Lam=1.63~\mathrm{GeV}$, $N_f=2$, and $e^2=4\pi/137.036$.
 }
  \label{fig:digamma-crossover}
\end{figure}

The different analytic structures at finite and vanishing temperature
can be seen directly from the subtracted Matsubara sum,
\begin{align}
  \mathcal D(\tau_I)
  &=
  -
  \sum_{n=1}^{\infty}
  \frac{\tau_I^2}{n\left(n^2+\tau_I^2\right)}
  \nonumber\\
  &=
  \sum_{\ell=1}^{\infty}
  (-1)^\ell\zeta(2\ell+1)\tau_I^{2\ell},
  \quad |\tau_I|<1.
  \label{eq:digamma-small}
\end{align}
At fixed \(T>0\), one has \(\tau_I=\order(e)\) in the weak-coupling
limit.  Equation~\eqref{eq:digamma-small} therefore only generates
integer powers of \(e^2\), starting with an additional factor of $e^2$
relative to the normalization $e^2\mE^4$. For $T\le\order(\mE)$,
the screening arguments are instead of order one and the nonstatic
contribution is of order $e^6$. By contrast, at fixed \(\mu\), the screening
argument grows parametrically as \(\tau_I\propto \mE/T\), and its cold
limit is governed by
\begin{align}
  \mathcal D(\tau_I)
  &\sim
  -\ln\tau_I-\gE
  -\sum_{\ell=1}^{\infty}
  \frac{(-1)^{\ell+1}B_{2\ell}}
       {2\ell\,\tau_I^{2\ell}}
  \nonumber\\
  =&
  -\ln\tau_I-\gE
  -\frac{1}{12\tau_I^2}
  -\frac{1}{120\tau_I^4}
  -\frac{1}{252\tau_I^6}
  +\order(\tau_I^{-8}),
  \label{eq:digamma-large}
\end{align}
where \(B_{2\ell}\) are the Bernoulli numbers.  The exact crossover and
the two asymptotic expansions are compared in
\cref{fig:digamma-crossover}.

We take the cold limit at fixed \(\mu\), for which
\begin{equation}
  \mE^2\mapsto\frac{N_fe^2\mu^2}{\pi^2},
  \qquad
  \mathcal I_T\mapsto
  2-2\ln\left(\frac{2\lvert\mu\rvert}{\Lam}\right).
  \label{eq:cold-inputs}
\end{equation}
In the same limit, the LO, one-loop power, and two-loop NLO
self-energies collected in \cref{app:selfenergies} reduce to the
zero-temperature expressions derived in
Ref.~\cite{Gorda:2022fci}. At strictly $T=0$, the $p_\mix^\naive$ contribution is homogeneous and scale-free and vanishes in
dimensional regularization. The zero-temperature result therefore follows
from the dimensionally regulated expression in
\cref{app:resummed-cold}, with the cold limit taken before the
expansion in $\eps$.
The surviving $1/\eps$ pole contributes to the finite part through its
product with the $\order(\eps)$ self-energies, screening mass, multiplicity,
and angular measure. These terms cancel from the subtracted result at
$T>0$ and must be retained at $T=0$.
At $T \to 0$ the even Matsubara sum becomes a continuous energy integral:
\begin{equation}
  T f(0)+2T\sum_{n=1}^{\infty}f(2\pi nT)
  \longrightarrow
  \int_{-\infty}^{\infty}
  \frac{\dd k_0}{2\pi}\,f(k_0).
  \label{eq:sum-to-integral}
\end{equation}
Thus, the static and nonstatic sectors reconstruct the corrected cold mixed
result of Ref.~\cite{Gorda:2022zyc} when this limit is taken in the expression. The already finite $\gHTL$ has the small-$T$ behavior
\begin{equation}
  \left.p_{\mix}\right|_{T>0,\,T\to0}
  =2\left.A_{-1}^{\mix}\right|_{T=0}
    \ln\frac{2\pi T}{\mE}
    +\order(T^0),
  \label{eq:finite-T-cold-log}
\end{equation}
at fixed $\mu\ne0$, $e$, and renormalization scale, with
$A_{-1}^{\mix}$ given by
\begin{equation}
  A_{-1}^{\mix}
  =-\frac{e^2\mE^4}{3072\pi^4}
  \left[
    N_f\left(4\mathcal I_T+\frac{7\pi^2}{12}-13\right)
    +3\frac{
      \mu^2(66-5\pi^2)+2\pi^2T^2(15-\pi^2)
    }{M}
  \right].
  \label{eq:complete-mixed-pole}
\end{equation}
Here $M\coloneqq3\mu^2+\pi^2T^2$. This logarithic IR divergence for small $T$ is compensated by the
naive hard contribution in the same scheme.

The logarithm in \cref{eq:digamma-large} is also responsible for the
nonanalytic coupling dependence of the cold result: in the unsubtracted regulated expression $p_\mix^\res$ it replaces the
Matsubara scale by the electric screening scale.  The resulting mixed
term \(2A_{-1}^{\mix}\ln(\Lamh/\mE)\) from \cref{eq:complete-mixed-pole} combines with the hard
contribution, canceling the factorization scale and leaving
\begin{align}
  \left.p_{\hard}+p_{\mix}\right|_{T=0,\,\ln}
  &=
  2\left.A_{-1}^{\mix}\right|_{T=0}
  \ln\left(\frac{Q_{\hard}}{\mE}\right)
  \nonumber\\
  &=
  2\left.A_{-1}^{\mix}\right|_{T=0}
  \left[-\ln e+\order(e^0)\right],
  \label{eq:cold-coupling-log}
\end{align}
where \(Q_{\hard}=\order(\lvert\mu\rvert)\).  Since
\(A_{-1}^{\mix}=\order(e^6)\), this produces the zero-temperature
order-\(e^6\ln e\) contribution.  It arises only through the
nonuniform cold limit; at every fixed \(T>0\),
\cref{eq:digamma-small} remains analytic in \(e^2\).
\subsection{Numerical results}
\label{sec:numerical}

For \(N_f=1\), we evaluate the mixed contribution along the
constant-hard-scale trajectory \cite{Kurkela:2016was}
\begin{equation}
  Q(T,\mu)
  =\sqrt{(2\mu)^2+\bigl(0.723\times4\pi T\bigr)^2}
  =Q_\star,
  \quad Q_\star\coloneqq2~\mathrm{GeV},
  \label{eq:numerical-trajectory}
\end{equation}
which connects the cold endpoint
\((T,\mu)=(0,1~\mathrm{GeV})\) to the hot endpoint
\((T,\mu)\simeq(220~\mathrm{MeV},0)\).  We set
\(\Lam=XQ_\star\), with \(X=1\) for the central result and
\(1/2\leq X\leq2\) for the scale variation, and evolve the coupling from
\(\alpha_e(Q_\star)^{-1}=137.036\).\footnote{We use one-loop running,
$\alpha_e(XQ_\star)^{-1}=137.036-[2N_f/(3\pi)]\ln X$.}
The finite-$T$ result is independent of $\Lamh$; for the separate
strict-$T=0$ evaluation we may set $\Lamh=XQ_\star$.
We normalize by the ideal pressure
\begin{equation}
  p_{\LO}
  =\frac{\pi^2T^4}{45}
  +N_f\left(
    \frac{7\pi^2T^4}{180}
    +\frac{\mu^2T^2}{6}
    +\frac{\mu^4}{12\pi^2}
  \right).
  \label{eq:ideal-pressure}
\end{equation}

\begin{figure}[t]
  \centering
  \includegraphics[width=0.98\linewidth]{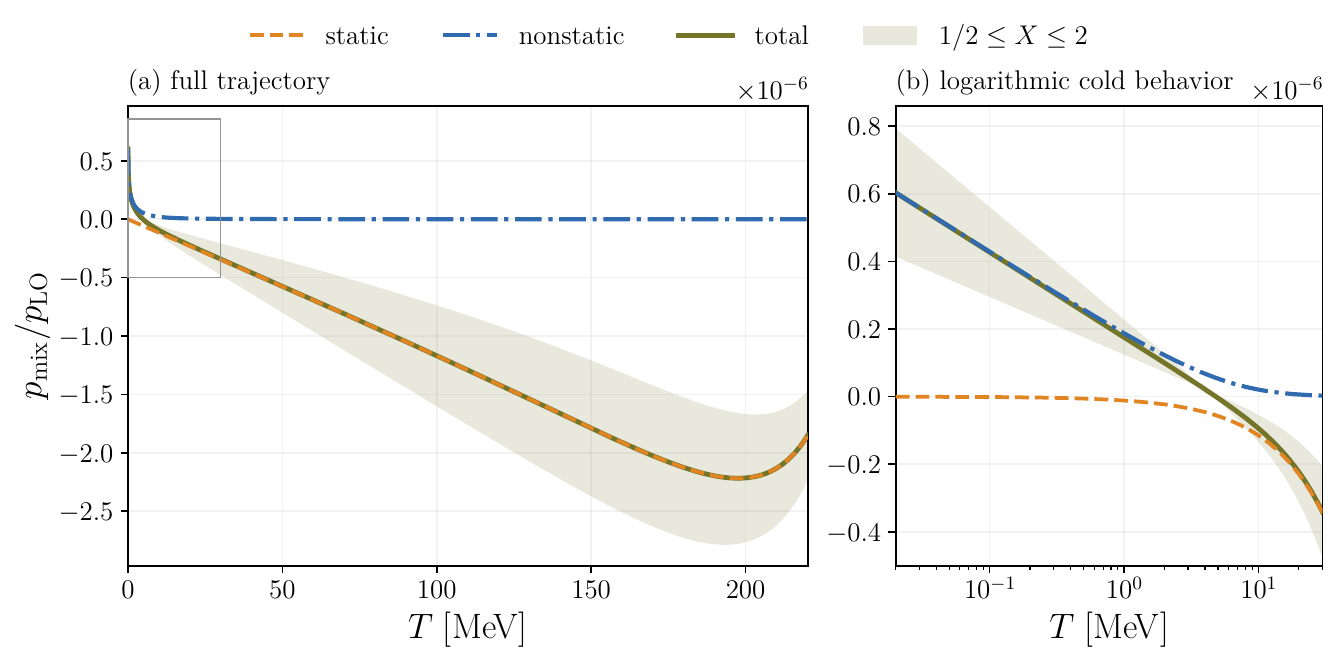}
  \caption{(Left) Mixed contribution at $T>0$, normalized by \cref{eq:ideal-pressure} along the trajectory \cref{eq:numerical-trajectory}. 
  The orange dashed and blue dot-dashed curves show the static and nonstatic contributions, respectively, and the olive curve their sum.
  The light-gray box  marks the region shown on the right.
  (Right) A zoom in to resolve the logarithmic small-$T$ behavior; the strict $T=0$ contribution is evaluated separately in \cref{app:resummed-cold}. 
  The shaded envelope corresponds to \(1/2\leq X\leq2\) in both cases.}
  \label{fig:mixed-pressure-trajectory}
\end{figure}
As shown in \cref{fig:mixed-pressure-trajectory}, the static contribution
vanishes at the cold endpoint, while the subtracted nonstatic sum has the
logarithmic small-$T$ behavior in \cref{eq:finite-T-cold-log}.
Away from this endpoint, the static mode supplies most of the temperature
dependence and dominates near \(\mu=0\), consistently with the \(\order(e^5)\)
result in \cref{eq:hot-e5-match}. The shaded band quantifies the hard
renormalization-scale dependence of the mixed contribution.

Numerical results for the self-energy formulas and the nonstatic $g_\HTL$
are provided in an accompanying Mathematica file, available in a Zenodo repository~\cite{QEDNumericalChecks}.
\section{Conclusions}
\label{sec:summary}

We have obtained the infrared contribution to the pressure
of hot and dense massless QED at general temperature and chemical
potential. The static contribution is given in closed form by
\cref{eq:static-final}, while the exact nonstatic Matsubara sum reduces to
the one-dimensional digamma integrals in \cref{eq:gHTL-explicit}.
At $\mu=0$, the static sector reproduces the complete known $\order(e^5)$
pressure. At fixed $T>0$ and $\mu$, the nonstatic contribution is analytic
in $e^2$ and starts at order $e^8$; for $T\le\order(\mE)$, it contributes
at order $e^6$.

In the finite-temperature subtraction scheme used here, the HTL overlap
cancels the factorization pole before the dimensional expansion. The
mixed contribution therefore requires only the $\eps^0$ self-energies
and is independent of $\Lamh$. The strict cold result instead follows
from the regulated, unsubtracted expression in
\cref{app:resummed-cold}, with the cold limit taken before expanding
in $\eps$. The dimensional coefficients then contribute through the
surviving pole, recovering the corrected zero-temperature mixed pressure. Combining
this result with the hard sector produces the $\order(e^6\ln e)$ term,
whereas the logarithmic small-$T$ dependence of the finite-temperature
mixed contribution cancels against the hard contribution in the same
scheme. These features distinguish the weak-coupling and cold limits.

The calculation of the mixed pressure of QCD should proceed similarly,
albeit with more complicated self-energies. The nonvanishing soft sector
also leads to a more complex infrared structure. Moreover, as in
Ref.~\cite{Kurkela:2016was}, one must add and subtract an additional
contribution from the static Matsubara modes, whose behavior is captured
by electrostatic QCD~\cite{Appelquist:1981vg,Braaten:1995cm,Braaten:1995jr}.
We leave a thorough investigation of these points and the full evaluation
of the QCD case to future work.

\section*{Acknowledgments}
The authors thank Pablo Navarrete for detailed comments on the manuscript and suggestions for improvement. They also thank Risto Paatelainen for suggesting this computation, and Pranav Sharma for reading a preliminary version of the manuscript and providing useful feedback.
\appendix
\crefalias{section}{appendix}
\section{Finite-\texorpdfstring{\(T,\mu\)}{T,mu} self-energies}
\label{app:selfenergies}

This appendix collects the Euclidean photon self-energies through
\(\order(\eps)\). Their $\eps^0$ parts determine the
mixed pressure at $T>0$, while the dimensional terms enter
the strict cold calculation in \cref{app:resummed-cold}.
They follow from the
soft-momentum expansion of Ref.~\cite{Gorda:2022fci}.

\subsection{Euclidean projection}

Let $\bigl(\Pi_X^{\mathrm E}\bigr)_{\mu\nu}(K_n)$ denote the Euclidean photon polarization tensor in the $X$ sector, evaluated at $K_n=(\omega_n,\vec{k})$; subscripts $00$ and $\mu\mu$ denote its temporal component and Euclidean trace.
For the power correction, $\bigl(\Pi_{\Pow}^{\mathrm E}\bigr)_{\mu\nu}$
denotes the coefficient of $K_n^2$, as in \cref{eq:selfenergy-expansion}.
After the continuation
\(k^0\to\ii\omega_n\), the transverse and longitudinal self-energies follow
from the radial--angular results of Ref.~\cite{Gorda:2022fci} through
\begin{align}
  \Pi_{\mathrm L}^{X}(K_n)
  &=\frac{K_n^2}{k^2}
    \bigl(\Pi_X^{\mathrm E}\bigr)_{00},\nonumber\\
  \Pi_{\mathrm T}^{X}(K_n)
  &=\frac{1}{d-1}
    \left[
      \bigl(\Pi_X^{\mathrm E}\bigr)_{\mu\mu}
      -\Pi_{\mathrm L}^{X}(K_n)
    \right],
  \label{eq:Euclidean-projectors}
\end{align}
Here $X\in\{\LO,\Pow,\NLO\}$.
\subsection{Thermal functions}
\label{app:thermal-functions}

For \(x>0\), we define the following two functions, which appear in the expressions for the self-energies
\begin{align}
    L_E(x)&\coloneqq\frac{\arctan x}{x}, \\
  H_E(x)
  &\coloneqq L_E(x)\left[
      2+\ln\left(\frac14+\frac{1}{4x^2}\right)
    \right]
  -\frac{\ii}{2x}\left[
      \Li_2\left(\frac{\ii+x}{\ii-x}\right)
      -\Li_2\left(\frac{\ii-x}{\ii+x}\right)
    \right],
  \label{eq:HE-full}
\end{align}
where $\Li_s$ is the standard polylogarithm function. 
The function $H_E$ is real for real \(x\).  
The required thermal derivatives of the polylogarithm functions are related to derivatives of the Hurwitz zeta function.
They are
\begin{align}
  \Phi_1(\mubar)
  &\coloneqq
    \Li^{(1)}_0(-\e^{2\pi\mubar})+\Li^{(1)}_0(-\e^{-2\pi\mubar})
  =-\ln(2\pi\e^{\gE})-\frac{\aleph(w)}{2},
  \label{eq:Phi1-definition}\\
  \mathcal L_2^{(1)}(\mubar)
  &\coloneqq
    \Li^{(1)}_2(-\e^{2\pi\mubar})+\Li^{(1)}_2(-\e^{-2\pi\mubar})
  \nonumber\\*
  &=-2\pi^2\left[
     \ln(2\pi\e^{\gE-1})\left(\mubar^2+\frac{1}{12}\right)
     -\aleph(1,w)\right],
  \label{eq:L2-definition}\\
  \Psi_1(\mubar)
  &\coloneqq
    \Li^{(1)}_1(-\e^{2\pi\mubar})-\Li^{(1)}_1(-\e^{-2\pi\mubar})
  =-2\pi\mubar\ln(2\pi\e^{\gE})-\ii\pi\aleph(0,w),
  \label{eq:Psi-definition}
\end{align}
where $\Li_m^{(n)}(z) \coloneqq \lim_{s \to m} \partial_s^n \Li_s(z)$.
Following Ref.~\cite{Vuorinen:2003fs}, we use $w\coloneqq\frac12-\ii\mubar$,
$\aleph(w)$ as defined below \cref{eq:Hd-eval}, and
\begin{equation}
 \aleph(n,w)\coloneqq\zeta'(-n,w)+(-1)^{n+1}\zeta'(-n,w^*),\qquad
 \zeta'(s,z)\coloneqq\partial_s\zeta(s,z),\quad n\in\mathbb Z_{\ge0}.
\end{equation}
We also define a new function 
\begin{equation}
    \beth(w)\coloneqq\gamma_1(w)+\gamma_1(w^*),
\end{equation}
which is related to
the generalized Stieltjes constants in the Laurent expansion
\begin{equation}
 \zeta(s,z)=\frac1{s-1}+\sum_{n=0}^{\infty}
 \frac{(-1)^n}{n!}\gamma_n(z)(s-1)^n,\qquad\gamma_0(z)=-\psi(z).
\end{equation}

The function \(\mathcal I_T\) is given in
\cref{eq:IT-result-explicit}.  Its \(\order(\eps)\) companion is
\begin{equation}
  \mathcal I_T^{[2]}
  =\mathcal I_T^2+\pi^2-\aleph(w)^2-4\beth(w).
  \label{eq:IT2-full}
\end{equation}
For later use, define the following expressions 
\begin{align}
  M&\coloneqq3\mu^2+\pi^2T^2,\quad
  S\coloneqq\mu^2+\pi^2T^2, \\
  \mathcal X(T,\mu)
  &\coloneqq
  -\frac{9S}{4M^2}\Phi_1(\mubar)
  +\frac{9T^2(M+4\pi^2T^2)}{2M^3}
    \mathcal L_2^{(1)}(\mubar)
  +\frac{9T^2}{M^2}\zeta'(2),
  \label{eq:X-master}\\
  \mathcal A(T,\mu)
  &\coloneqq
  \mu^2+4\mu T\Psi_1(\mubar)
  -\frac{24\mu^2T^2}{M}\mathcal L_2^{(1)}(\mubar).
  \label{eq:Amu-master}
\end{align}
The dimensional continuation of the screening parameter in the radial
representation of Ref.~\cite{Gorda:2022fci} satisfies
\begin{align}
  \frac{\mEd^2}{\mE^2}
  &=1+\eps\,\delta_E(T,\mu)+\order(\eps^2),
  \nonumber\\
  \delta_E(T,\mu)
  &=2\ln\frac{\mathrm{e}^{\gE}\Lam}{2T}-1
    +\frac{12T^2}{M}\mathcal L_2^{(1)}(\mubar).
  \label{eq:screening-dimensional-factor}
\end{align}
\subsection{T, L components}
In the LO and NLO self-energies below, we factor out
\(\mEd^{2}\) and leave this factor unexpanded while expanding
the coefficients through \(\order(\eps)\). Its value at \(d=3\)
is \(\mE^2\) in \cref{eq:screening-mass}, with their relation
given in \cref{eq:screening-dimensional-factor}. The leading-order T and L
components of the renormalized self-energies are
\begin{align}
  \Pi_{\mathrm T}^{\LO}(x)
  &=\frac{\mEd^2}{2}\Bigg\{
    (1+\eps)\left[\left(1+x^{-2}\right)L_E-x^{-2}\right]
    -\eps\left(1+x^{-2}\right)H_E
  \Bigg\}+\order(\eps^2),
  \label{eq:LO-T-full}\\
  \Pi_{\mathrm L}^{\LO}(x)
  &=\mEd^2\Bigg\{
    \left(1+x^{-2}\right)(1-L_E)
    +\eps\left(1+x^{-2}\right)H_E
  \Bigg\}+\order(\eps^2).
  \label{eq:LO-L-full}
\end{align}
After cancellation of the vacuum pole against the coupling counterterm, the renormalized one-loop power corrections are
\begin{align}
  \Pi_{\mathrm T}^{\Pow}(x)
  &=\frac{N_fe^2}{24\pi^2}\Bigg\{
    -\left[
      1-2\mathcal I_T
      +\left(3-x^{-2}\right)(1-L_E)
    \right]
  +\eps\Bigg[
      4-\frac{\pi^2}{2}-\mathcal I_T+\mathcal I_T^{[2]}
  \nonumber\\*[-0.2em]
  &\qquad
       -\left(3-x^{-2}\right)H_E
       -\left[3\mathcal I_T+(2-\mathcal I_T)x^{-2}\right](1-L_E)
    \Bigg]
  \Bigg\}+\order(\eps^2),
  \label{eq:Pow-T-full}\\
  \Pi_{\mathrm L}^{\Pow}(x)
  &=\frac{N_fe^2}{12\pi^2}\Bigg\{
    \mathcal I_T
    -\left(3+x^{-2}\right)(1-L_E)
 +\eps\Bigg[
      1-\frac{\pi^2}{4}+\frac{\mathcal I_T^{[2]}}{2}
      -\left(3+x^{-2}\right)H_E
  \nonumber\\*[-0.2em]
  &\qquad
      -\left\{
        2\mathcal I_T
        -(3-\mathcal I_T)\left(1+x^{-2}\right)
      \right\}(1-L_E)
    \Bigg]
  \Bigg\}+\order(\eps^2).
  \label{eq:Pow-L-full}
\end{align}
The renormalized result is obtained by subtracting the channel-independent pole
$\frac{N_fe^2}{12\pi^2\epsUV}$.
Using the abbreviations in
\cref{eq:X-master,eq:Amu-master}, the renormalized two-loop NLO self-energies
take the compact form
\begin{align}
  \Pi_{\mathrm T}^{\NLO}(x)
  &=e^2\mEd^2\Bigg\{
    -\frac{3S}{8\pi^2M}L_E
 +\eps\Bigl[
      \frac{3S}{8\pi^2M}(H_E-3L_E)
      +\frac{M}{3\pi^2}L_E \mathcal X(T,\mu)
  \nonumber\\*[-0.2em]
  &\qquad\quad
      -\frac{3S}{8\pi^2M}\delta_E(T,\mu)L_E
    \Bigr]
  \Bigg\}+\order(\eps^2),
  \label{eq:NLO-T-full}\\
  \Pi_{\mathrm L}^{\NLO}(x)
  &=e^2\mEd^2\Bigg\{
    -\frac{3}{8\pi^2M}\left[
      S+2\mu^2\left(1+x^{-2}\right)(1-L_E)^2
    \right]-\eps\frac{3S}{8\pi^2M}(1+2L_E)
  \nonumber\\*[-0.2em]
  &\quad+\eps\frac{M}{3\pi^2}\mathcal X(T,\mu)-\eps\frac{3(1+x^2)(L_E-1)}{4\pi^2x^2M}
    \left[
      \mathcal A(T,\mu)(L_E-1)-2\mu^2H_E
    \right]
  \nonumber\\*[-0.2em]
  &\quad-\eps\frac{3\delta_E(T,\mu)}{8\pi^2M}
    \left[
      S+2\mu^2\left(1+x^{-2}\right)(1-L_E)^2
    \right]
  \Bigg\}+\order(\eps^2).
  \label{eq:NLO-L-full}
\end{align}
\section{Static limits and common pole residue}
\label{app:checks}

The vacuum pole of the one-loop power correction cancels mode by mode
against the coupling counterterm of the lower-order ring pressure.
As the contribution from the zero mode is finite in our subtraction scheme, in this section, we drop all the unneeded $\order(\eps)$ terms.

At $K=(0,\vec{k})$, the static limits needed in
\cref{eq:static-before-integral} are
\begin{align}
  \Pi_{\mathrm T}^{\LO}(0,\vec{k})&=0,
  \quad \Pi_{\mathrm L}^{\LO}(0,\vec{k})=\mE^2,
  \nonumber\\
  \Pi_{\mathrm L}^{\NLO}(0,\vec{k})
  &=-\frac{3e^2\mE^2}{8\pi^2},\quad
  \Pi_{\mathrm L}^{\Pow}(0,\vec{k})
  =\frac{N_fe^2}{12\pi^2}(\mathcal I_T-3).
  \label{eq:static-checks}
\end{align}
The two angular reductions determining the common pole of resummed mixed term $p_\mix^\res$ and its naive counterpart $p_\mix^\naive$ are
\begin{align}
  \int_0^\infty
  \frac{\dd x\,x^2}{(1+x^2)^2}
  \left.
  \sum_I g_I\big[-\Pi_I^{\Pow}(\Pi_I^{\LO})^2\big]
  \right|_{\eps=0}
  &=-\frac{N_fe^2\mE^4}{384\pi}
  \left(4\mathcal I_T+\frac{7\pi^2}{12}-13\right),
  \label{eq:power-angular-check}
\end{align}
\begin{equation}
\begin{aligned}
  \int_0^\infty
  \frac{\dd x\,x^2}{(1+x^2)^2}
  \left.
  \sum_I g_I\Pi_I^{\NLO}\Pi_I^{\LO}
  \right|_{\eps=0}
   &=-\frac{e^2\mE^4}{128\pi M}
  \left[
    \mu^2(66-5\pi^2)+2\pi^2T^2(15-\pi^2)
  \right].
\end{aligned}
  \label{eq:NLO-angular-check}
\end{equation}
Multiplying their sum by the pole prefactor $1/(8\pi^3)$ gives the
common residue $A_{-1}^{\mix}$ in \cref{eq:complete-mixed-pole} for
$p_{\mix}^{\res}$ and $p_{\mix}^\naive$. This residue cancels
from $p_{\mix}$ at $T>0$ and remains in the strict $T=0$ mixed contribution.

\section{Finite-temperature nonstatic coefficient}
\label{app:finite-masters}

At $T>0$, the difference of the resummed and naive sums in
\cref{eq:Hd-eval} permits evaluation of the self-energy kernels at
$\eps=0$. The angular weight is
$w(x)\coloneqq x^2/(1+x^2)^2$, with polarization multiplicities
$g_{\mathrm T}=2$ and $g_{\mathrm L}=1$.
The screening arguments in \cref{eq:screening-argument} reduce to
\begin{equation}
 \tau_{\mathrm T,0}^2(x)
 =\frac{\mE^2}{8\pi^2T^2x^2}
   \left[L_E-\frac{1}{1+x^2}\right],\qquad
 \tau_{\mathrm L,0}^2(x)
 =\frac{\mE^2}{4\pi^2T^2x^2}(1-L_E).
 \label{eq:finite-screening-arguments}
\end{equation}
Integrating the constant term in \cref{eq:Hd-eval} using
\cref{eq:power-angular-check,eq:NLO-angular-check} and substituting the
self-energy kernels gives
\begin{align}
 \gHTL={}&\frac{\gE}{6}\Biggl[
   N_f\left(4\mathcal I_T+\frac{7\pi^2}{12}-13\right)
   +3\frac{\mu^2(66-5\pi^2)+2\pi^2T^2(15-\pi^2)}{M}
 \Biggr]
 \nonumber\\*
 &+\frac{4}{3\pi}\int_0^\infty\frac{\dd x}{x^2}\Biggl\{
   N_f\left[2\mathcal I_T-1-\left(3-\frac1{x^2}\right)(1-L_E)\right]
   \left[L_E-\frac1{1+x^2}\right]^2
 \nonumber\\*
 &\hspace{21mm}
   +\frac{18Sx^2}{M(1+x^2)}L_E
   \left[L_E-\frac1{1+x^2}\right]
 \Biggr\}\operatorname{Re}\psi(1+\ii\tau_{\mathrm T,0})
 \nonumber\\*
 &+\frac{8}{3\pi}\int_0^\infty\frac{\dd x}{x^2}(1-L_E)\Biggl\{
   2N_f\left[\mathcal I_T-\left(3+\frac1{x^2}\right)(1-L_E)\right]
   (1-L_E)
 \nonumber\\*
 &\hspace{21mm}
   +\frac{9}{M}\Bigl[
     \frac{Sx^2}{1+x^2}+2\mu^2(1-L_E)^2
   \Bigr]
 \Biggr\}\operatorname{Re}\psi(1+\ii\tau_{\mathrm L,0}).
 \label{eq:gHTL-explicit}
\end{align}
\section{Resummed mixed contribution and cold limit}
\label{app:resummed-cold}

We give the dimensionally regulated resummed mixed contribution at
general $T$ and $\mu$, and identify its naive counterpart in the
convention of \cref{eq:hard-overlap-matching}. Keeping these terms
separate makes explicit the dimensional coefficients needed for the
strict cold result.

For the nonzero Matsubara modes, the resummed term contains $H_d(\tau_I)$,
while the naive term contains $H_d(0)$. At fixed $T>0$,
$H_{3-2\eps}(\tau)=1/(2\eps)-\operatorname{Re}\psi(1+\ii\tau)
+\order(\eps)$. We therefore define their finite coefficients by
\begin{align}
  \left.p_{\mix}^{\res}\right|_{n\ne0}
  &\eqcolon\frac{A_{-1}^{\mix}}{\eps}
    +\frac{e^2\mE^4}{(4\pi)^4}\,\gres+\order(\eps),
  \label{eq:D-res-normalization}\\
  \left.p_{\mix}^{\naive}\right|_{n\ne0}
  &\eqcolon\frac{A_{-1}^{\mix}}{\eps}
    +\frac{e^2\mE^4}{(4\pi)^4}\,\gnaive+\order(\eps),
  \label{eq:D-naive-normalization}
\end{align}
where the common residue is given in \cref{eq:complete-mixed-pole}.
Their difference gives the finite-temperature coefficient in
\cref{eq:gHTL-definition},
\begin{equation}
  \gHTL=\gres-\gnaive.
  \label{eq:D-g-res-naive}
\end{equation}
The static naive term vanishes as a scaleless integral, and the static
resummed term is given by \cref{eq:static-final}.

To evaluate the finite coefficients, we use the dimensionless
self-energy kernels
\begin{align}
  \frac{\Pi_I^{\LO}}{\mEd^2}
  &\eqcolon\pi_{I,0}+\eps\pi_{I,1}+\order(\eps^2),\nonumber\\
  \frac{24\pi^2\Pi_I^{\Pow}}{N_fe^2}
  &\eqcolon P_{I,0}+\eps P_{I,1}+\order(\eps^2).
  \label{eq:D-selfenergy-kernels}
\end{align}
and $B_{I,0}\coloneqq[\Pi_I^{\NLO}/(e^2\mEd^2)]_{\eps=0}$.
Their explicit expressions are given in \cref{app:selfenergies}; the
weight $w(x)$ and screening arguments $\tau_{I,0}(x)$ are defined in
\cref{app:finite-masters}. We separate the thermal functions from the
power kernels by setting
$\widehat P_{I,j}\coloneqq
\left.P_{I,j}\right|_{\mathcal I_T=\mathcal I_T^{[2]}=0}$ and define
\begin{align}
  \mathcal K_{\Pow,0}
  &\coloneqq2\widehat P_{\mathrm T,0}\pi_{\mathrm T,0}^2
    +\widehat P_{\mathrm L,0}\pi_{\mathrm L,0}^2,\nonumber \\
  \mathcal K_{\Pow,1}
  &\coloneqq2\widehat P_{\mathrm T,1}\pi_{\mathrm T,0}^2
    +4\widehat P_{\mathrm T,0}\pi_{\mathrm T,0}\pi_{\mathrm T,1}
    -2\widehat P_{\mathrm T,0}\pi_{\mathrm T,0}^2+\widehat P_{\mathrm L,1}\pi_{\mathrm L,0}^2
    +2\widehat P_{\mathrm L,0}\pi_{\mathrm L,0}\pi_{\mathrm L,1}.
\end{align}
The $-2\widehat P_{\mathrm T,0}\pi_{\mathrm T,0}^2$ term comes from
$g_{\mathrm T}=2-2\eps$. The six dimensionless angular integrals
needed to reduce the finite coefficient are
\begin{align}
  \chi_{\mathrm L}
  &\coloneqq2\int_0^\infty\dd x\,w(x)\pi_{\mathrm L,0}^3
   =\frac{\pi}{48}(28-\pi^2-16\ln2),
  \label{eq:D-chiL}\\
  \kappa_{\Pow}
  &\coloneqq-\frac12\int_0^\infty\dd x\,w(x)
    [\mathcal K_{\Pow,1}-2\ln x\,\mathcal K_{\Pow,0}] \nonumber\\*
  &=\frac{\pi}{384}
    [-1332+142\pi^2-234\pi^2\ln2+1011\zeta(3)],
  \label{eq:D-kappaPow}\\
  \Delta_S
  &\coloneqq\int_0^\infty\dd x\,w(x)\ln x
    (2L_E\pi_{\mathrm T,0}+\pi_{\mathrm L,0})=\frac{\pi}{16}[20\ln2-7\zeta(3)-2],
  \label{eq:D-DeltaS}\\
  \Delta_\mu
  &\coloneqq2\int_0^\infty\dd x\,w(x)\ln x
    (1+x^{-2})(1-L_E)^2\pi_{\mathrm L,0}
  =\frac{\pi}{64}[12-\pi^2+96\ln2-42\zeta(3)],
  \label{eq:D-DeltaMu}\\
  c_S
  &\coloneqq\frac{3}{8\pi^2}\int_0^\infty\dd x\,w(x)
    \Bigl[2(H_E-2L_E)\pi_{\mathrm T,0}-2L_E\pi_{\mathrm T,1}
    -(1+2L_E)\pi_{\mathrm L,0}-\pi_{\mathrm L,1}\Bigr]
  \nonumber\\*
  &=\frac{-12+\pi^2-60\ln2+27\zeta(3)}{64\pi},
  \label{eq:D-cS}\\
  c_\mu
  &\coloneqq\frac{3}{4\pi^2}\int_0^\infty\dd x\,w(x)(1+x^{-2})
    [2(L_E-1)H_E\pi_{\mathrm L,0}-(1-L_E)^2\pi_{\mathrm L,1}]
  \nonumber\\*
  &=\frac{9}{256\pi}(4-\pi^2-32\ln2+4\pi^2\ln2).
  \label{eq:D-cMu}
\end{align}
Here and below, the $x$ dependence of the kernels is suppressed.
The remaining polarization sum uses $g_{\mathrm T,0}=2$ and
$g_{\mathrm L,0}=1$.

With these reductions, the finite coefficient of the resummed term is
\begin{align}
  \gres
  ={}&-\frac{N_f}{6}
  \left[1+\ln\frac{\Lamh}{4\pi T}+\delta_E(T,\mu)\right]
  \left(4\mathcal I_T+\frac{7\pi^2}{12}-13\right)\nonumber\\*
  &-\frac{1}{2M}
  \left[1+\ln\frac{\Lamh}{4\pi T}+\frac32\delta_E(T,\mu)\right]
  \left[\mu^2(66-5\pi^2)+2\pi^2T^2(15-\pi^2)\right]\nonumber\\*
  &+\frac{8N_f}{3\pi}\Biggl\{
    \kappa_{\Pow}
    +\left[\frac{\pi}{384}(364+128\ln2-31\pi^2)
      +\chi_{\mathrm L}\right]\mathcal I_T
    -\frac{\pi}{16}\mathcal I_T^{[2]}
  \Biggr\}\nonumber\\*
  &+\frac{24}{\pi M}\left(S\Delta_S+\mu^2\Delta_\mu\right)
  \nonumber\\*
  &+32\pi\left[
    \frac{S}{M}c_S+\frac{\mu^2}{M}c_\mu
    +\frac{M\mathcal X(T,\mu)(15-\pi^2)}{72\pi}
    -\frac{3\mathcal A(T,\mu)(12-\pi^2)}{128\pi M}
  \right]\nonumber\\*
  &+\int_0^\infty\dd x\,w(x)\sum_I g_{I,0}
  \left[\frac{8N_f}{3\pi}P_{I,0}\pi_{I,0}^2
    -64\pi B_{I,0}\pi_{I,0}\right]
  \operatorname{Re}\psi(1+\ii\tau_{I,0}(x)).
  \label{eq:D-g-res-reduced}
\end{align}
The functions $M$, $S$, $\delta_E$, $\mathcal X$, and $\mathcal A$ are
defined in \cref{app:thermal-functions}. The analytic terms include the
finite contributions from the pole multiplying the $\order(\eps)$
self-energies, screening mass, polarization multiplicity, and angular
measure. Only the digamma integral remains numerical.

The naive coefficient $\gnaive$ follows from the same expression by
setting only the screening arguments in the digamma factors to zero,
so that $\operatorname{Re}\psi(1+\ii\tau_{I,0})\mapsto-\gE$.
All analytic terms are common to the two coefficients and cancel in
\cref{eq:D-g-res-naive}. The remaining difference is precisely
\cref{eq:gHTL-explicit}, or equivalently the angular representation
\cref{eq:gHTL-angular-area,eq:Gphi-definition}.

At strictly $T=0$, the naive contribution is scaleless and the static
contribution vanishes. The mixed pressure is therefore given by the
resummed term. To take this limit, we keep the dimensional regulator
fixed and use
\begin{equation}
  H_{3-2\eps}(\tau)\sim
  \int_0^\infty\dd n\,\frac{n^{1-2\eps}}{n^2+\tau^2}
  =\frac{\pi}{2\sin(\pi\eps)}\,\tau^{-2\eps},
  \qquad0<\operatorname{Re}\eps<1.
  \label{eq:D-Hd-cold-regulated}
\end{equation}
Together with the $T^{-2\eps}$ prefactor in
\cref{eq:nonstatic-master}, this reconstructs the continuous energy
integral. At fixed $\mu\ne0$, the radial functions satisfy
\begin{align}
  \mathcal I_T,\delta_E&\mapsto
    \mathcal I_0\coloneqq2+2\ln\frac{\Lam}{2|\mu|},
  &\mathcal I_T^{[2]}&\mapsto\mathcal I_0^2,  &
  \mathcal X&\mapsto\frac{3}{8\mu^2},
  &\mathcal A&\mapsto-\mu^2.
  \label{eq:D-cold-radial-functions}
\end{align}
Expanding in $\eps$ afterward gives
\begin{equation}
  \left.p_{\mix}\right|_{T=0}
  =\frac{\left.A_{-1}^{\mix}\right|_{T=0}}{\eps}
    +\frac{e^2\mE^4}{(4\pi)^4}\,g_{\HTL}^{(0)}
    +\order(\eps),
  \label{eq:D-cold-pressure}
\end{equation}
with $\mE^2=N_fe^2\mu^2/\pi^2$.
The finite coefficient $g_{\HTL}^{(0)}$ is obtained from
\cref{eq:D-g-res-reduced} using \cref{eq:D-cold-radial-functions},
$M=3\mu^2$, and $S=\mu^2$. Before taking the limit, the temperature
logarithms must be combined with the large-argument digamma terms.
Using \cref{eq:power-angular-check,eq:NLO-angular-check}, the relevant
combination becomes
\begin{equation}
  \ln\frac{\Lamh}{4\pi T}
    -\operatorname{Re}\psi(1+\ii\tau_{I,0})
  \longrightarrow
  \ln\frac{\Lamh}{2\mE}
    -\frac12\ln\frac{\pi_{I,0}}{1+x^2}.
  \label{eq:D-cold-log-kernel}
\end{equation}
This recovers the cold mixed pressure of Ref.~\cite{Gorda:2022zyc}, with
the finite NLO contribution corrected as described in
\cref{sec:limiting-cases}. 
\bibliographystyle{JHEP}
\bibliography{refs}

\end{document}